\documentclass[twocolumn]{article}
\usepackage[a4paper, total={180mm, 243mm}]{geometry}
\usepackage[T1]{fontenc}
\usepackage{lipsum}
\usepackage{graphicx}
\usepackage{siunitx}
\usepackage{upgreek}
\usepackage{titling}
\usepackage{circledsteps}
\usepackage{subcaption}
\usepackage[version=4]{mhchem}
\usepackage{amsmath, amssymb}
\usepackage{soul}

\definecolor{authorcolor}{cmyk}{0.8,0.6,0,0.3}

\DeclareCaptionLabelSeparator{custom}{\textbar}
\DeclareCaptionFormat{custom}{\textsf{\textbf{#1 #2} \small #3}}
\graphicspath{{./figures}} 

\makeatletter
\newcommand{\showfontsize}{\f@size{} pt}
\makeatother

\usepackage[
    backend=biber,
    style=nature,
    url=false,
    isbn=false,
    date=year,
    maxnames=99
]{biblatex}
\let\cite\supercite

\newcommand{\reffig}[2]{Fig.~\ref{#1}{\textbf{#2}}}

\newcommand\identity{1\kern-0.25em\text{l}}

\title{Attosecond-level synchronisation of chip-integrated oscillators}

\author{Alexander~E.~Ulanov$^{1}$,
        Bastian Ruhnke$^{1}$,
        Theia El Sharkawy$^{2}$,\\
        Thibault Wildi$^{1}$,
        Kemal \c{S}afak$^{2}$,
        Franz Kärtner$^{1,2,3}$,
        Tobias Herr$^{1,3,*}$
}
\date{%
    \small $^1$Deutsches Elektronen-Synchrotron DESY, Notkestr. 85, 22607 Hamburg, Germany \\
    \small $^2$Cycle GmbH, Luruper Hauptstr. 1, 22547 Hamburg, Germany \\
    \small $^3$Physics Department, University of Hamburg UHH, Luruper Chaussee 149, 22761 Hamburg, Germany\\
    \small $^{*}$tobias.herr@desy.de
}

\begin{document}

\begin{refsection}
\maketitle

\textbf{
Synchronised laser oscillators are essential for probing the fastest processes in chemistry, materials science, and biology down to atto-second timescales.
Tight synchronisation is also crucial at scientific facilities such as free-electron lasers or radio-telescopes, and increasingly relevant to communication and information technologies in multi-node networks.
Current synchronisation approaches based on mode-locked lasers achieve the required performance, but their complexity, cost, and size hinder deployment in multi-node networks.
Here, we demonstrate attosecond-level synchronisation between chip-integrated microresonator soliton oscillators operating at either 25 or 300 GHz pulse repetition rate. For synchronisation, each oscillator receives over fibre a pair of continuous-wave lasers as a two-tone timing reference. The lasers power the microcombs and Kerr-nonlinear synchronisation results in integrated relative timing jitter below 400~as (1~kHz to 1~MHz), without any active stabilisation. This approach enables scalable precision timing for large facilities, data centres, disaggregated computing, navigation, and quantum networks; ultimately, it may lead to chip-integrated attosecond photonics.
}

The fastest processes in chemistry, materials science, and biology are governed by dynamics of bound electrons and occur on an attosecond time scale\cite{hentschel2001AttosecondMetrology, drescher2002TimeresolvedAtomicInnershell,
calegari2014UltrafastElectronDynamics, li2024AttosecondpumpAttosecondprobeXray}. Free-electron X-ray lasers \cite{emma2010FirstLasingOperation,allaria2012HighlyCoherentStable, decking2020MHzrepetitionrateHardXray, prat2023XrayFreeelectronLaser} (XFELs) and high-intensity laser beam lines\cite{kuhn2017ELIALPSFacilityNext} promise access to this regime, combined with sub-atomic spatial resolution. To reach this goal, facility-wide synchronisation on femto- or even attosecond level must be achieved\cite{schulz2015FemtosecondAllopticalSynchronization, hartmann2018AttosecondTimeEnergy, xin2018UltrapreciseTimingSynchronization, 
sato2020FemtosecondTimingSynchronization}. Moreover, emerging applications in wireless communication~\cite{nagatsuma2016AdvancesTerahertzCommunications}, data centers~\cite{clark2020SynchronousSubnanosecondClock, valley2007PhotonicAnalogtodigitalConverters}, and quantum key distribution~\cite{clivati2022CoherentPhaseTransfer} demand increasingly tight synchronisation of multi-node networks.

The most advanced synchronisation techniques rely on optical methods~\cite{foreman2007RemoteTransferUltrastable,xin2018UltrapreciseTimingSynchronization, caldwell2025HighprecisionOpticalTime}, exploiting low-loss fiber or free-space links to distribute timing references over long distances~\cite{wilcox2009StableTransmissionRadio,hudson2006SynchronizationModelockedFemtosecond,coddington2007CoherentOpticalLink}. Optical timing networks typically comprise multiple low-noise local oscillators phase-locked to a common reference delivered via fiber or free-space links.
Environmental perturbations introduce link-induced timing noise, causing accumulated timing error. This is mitigated through active link stabilisation, oscillator steering, or digital signal processing. Increasing feedback bandwidth reduces in-band timing jitter but raises system complexity and power consumption, and is fundamentally limited by actuator dynamics (e.g. approx. 1~kHz for fiber stretchers) and, in two-way schemes, by the link round-trip delay. 
Thus, a crucial metric is the integrated relative timing jitter of the link and the oscillators above the feedback bandwidth, defining the ultimate synchronisation performance.

Mode-locked lasers set the benchmark as low-noise oscillators offering, in particular, low timing jitter above typical link tracking feedback bandwidth of 1~kHz. However, they require active locking loops and are usually table-top systems, which complicates their use in multi-node networks. As a complementary technology, microcombs provide compact, low-cost, and scalable photonic oscillators, that can be integrated on photonic chips and potentially deployed in large networks\cite{kippenberg2018DissipativeKerrSolitons, pasquazi2018MicrocombsNovelGeneration, gaeta2019PhotonicchipbasedFrequencyCombs}. 
They are generated via nonlinear optical frequency conversion in a microresonator pumped by a continuous-wave (cw) laser~\cite{delhaye2007OpticalFrequencyComb}, enabling femtosecond pulses~\cite{herr2014TemporalSolitonsOptical} with repetition rates from 10~GHz to 1~THz, and low-noise signal synthesis ~\cite{lucas2020UltralownoisePhotonicMicrowave, kudelin2024PhotonicChipbasedLownoise, sun2024IntegratedOpticalFrequency, zang2025UniversalElectronicSynthesis}.
When a second (usually weaker) cw laser is injected into the microresonator, \textit{Kerr-nonlinear synchronisation} locks the repetition rate to an integer fraction of the frequency difference between the cw lasers\cite{jang2015TemporalTweezingLight, taheri2017OpticalLatticeTrap, weng2019SpectralPurificationMicrowave, brasch2019NonlinearFilteringOptical, moille2023KerrinducedSynchronizationCavity, wildi2023SidebandInjectionLocking, lei2024SelfinjectionlockedOpticalParametric}. Recently, this enabled the generation of ultra-pure microwaves\cite{sun2025MicrocavityKerrOptical, egbert2026AttosecondtimingMillimeterWaves} and also underpins synchronisation between microcombs or optical parametric oscillators within a microresonator\cite{yang2017CounterpropagatingSolitonsMicroresonators, zhang2020SpectralExtensionSynchronization, zhao2024AllopticalFrequencyDivision} and between microresonators \cite{jang2018SynchronizationCoupledOptical, kim2021SynchronizationNonsolitonicKerr}, including long-distance coherence transfer\cite{geng2022CoherentOpticalCommunications}.

\begin{figure*}[!t]
  \centering
  \includegraphics[width=\textwidth]{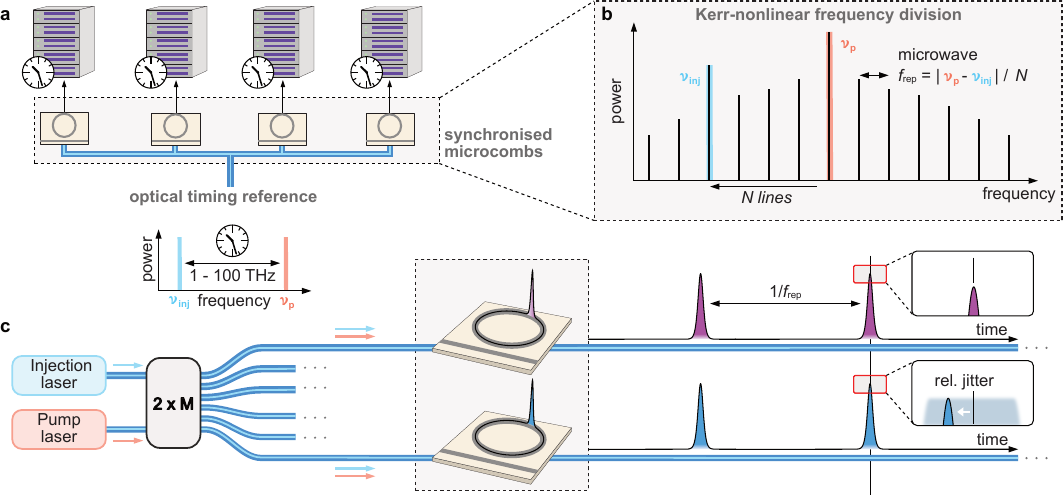}
  \caption{
    \textbf{Concept}. 
    \textbf{a}, Two laser frequencies ($\nu_\mathrm{p}$ and $\nu_\mathrm{inj}$) define an optical timing reference that is distributed over fiber to different nodes in a timing network, where synchronised microcombs provide local precision timing. \textbf{b}, Both laser frequencies $\nu_\mathrm{p}$ and $\nu_\mathrm{inj}$ are part of the microcombs spectrum, whose discrete spectral components are spaced by $f_\mathrm{rep}= |\nu_\mathrm{p}-\nu_\mathrm{inj}|/N$, where N is an integer. \textbf{c}, The two lasers with frequencies $\nu_\mathrm{p}$ and $\nu_\mathrm{inj}$ are combined and can then be distributed to $M$ different microcombs via a $2\times M$ fiber coupler. Both microcombs will emit pulses with the repetition rate $f_\mathrm{rep}$, however, uncorrelated noise in both systems will lead to timing jitter. 
    }
  \label{fig:concept}
\end{figure*}

Here, we demonstrate precision synchronisation of two microcombs on separate photonic chips. We achieve this via Kerr-nonlinear synchronisation based on two cw lasers that are delivered to the microcombs via optical fiber. The frequency difference between both lasers defines a two-tone timing reference to which the distributed microcombs synchronise without any active feedback loops (\reffig{fig:concept}{a}). In a \textit{first} experiment, we separate two 300~GHz microcombs by effectively 100~m of fiber (2x50~m). Using a frequency domain technique, we reveal an integrated relative timing jitter on the attosecond-level (1~kHz to 1 ~MHz), well below the microresonators' fundamental thermorefractive noise level. In a \textit{second} experiment, we synchronise two 25~GHz microcombs, and utilize an independent measurement technique -- balanced optical cross-correlation (BOC) -- to demonstrate and confirm the low relative timing jitter. Together, these experiments open opportunities for precision synchronisation in large networks and chip-integrated attosecond science.

\begin{figure*}[!t]
  \centering
  \includegraphics[width=\textwidth]{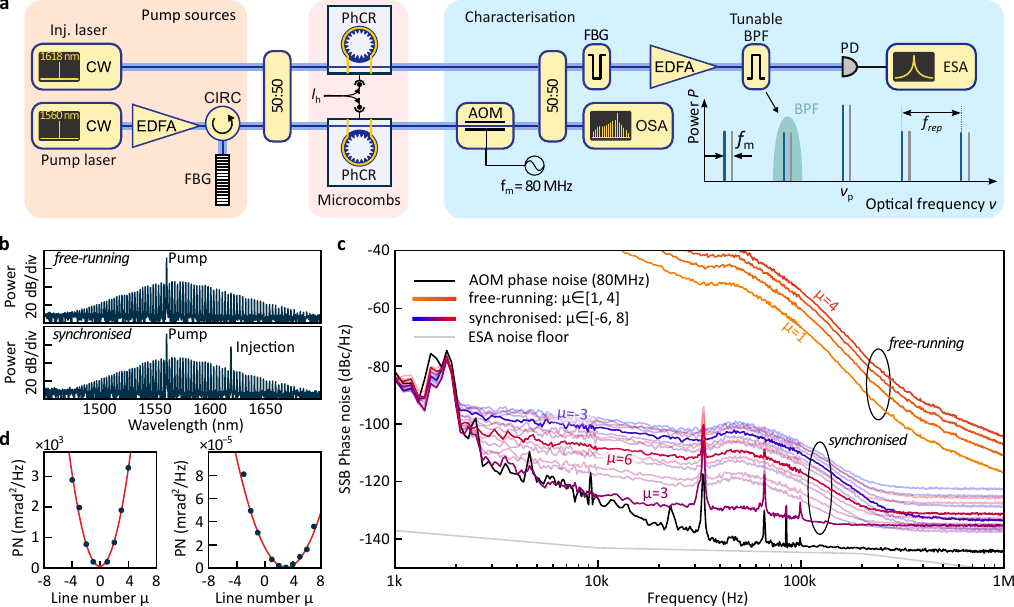}
  \caption{
    \textbf{Experimental approach to measuring the relative timing jitter of 300~GHz microcombs.}
    \textbf{a}, Schematic of the experimental setup. Two cw lasers (pump and injection) are combined on a 3 dB coupler and used to pump two identical photonic crystal resonators (PhCRs). An acousto-optic modulator (AOM) shifts the output of one comb by 80~MHz, after which both combs are recombined at a fiber coupler and amplified using an erbium-doped fiber amplifier (EDFA). A tunable bandpass filter (BPF) is used to select a single pair of comb lines (see inset). PD - photodiode, ESA - electrical spectrum analyser, FBG - fiber Bragg grating.
    \textbf{b}, Optical spectrum of two independent microcombs in the free-running (top, no injection) and synchronised (bottom, with injection) regimes.
    \textbf{c}, Single-sideband (SSB) phase-noise power spectral density (PSD) of the beatnotes between pairs of comb lines with index $\mu$, measured in the free-running and synchronised states. The measurement is limited by the phase-noise of the AOM, which is retrieved from the beating between shifted and unshifted pump (black) and the ESA noise floor (gray). 
    At high frequencies, the noise floor is dominated by amplified spontaneous emission.
    \textbf{d}, Phase-noise (PN) PSD at 10~kHz analysis frequency, extracted from the data in \textbf{c}, in the free-running (left) and synchronised (right) states. PN is shown using linear scaling ($\mathrm{mrad^2/Hz}$). 
    }
  \label{fig:sync}
\end{figure*}

\subsection*{Results}
The approach for synchronising the microcombs over fiber is as follows: A main pump laser of frequency $\nu_\mathrm{p}$ generates the microcomb, which in the spectral domain contains discrete and equidistant frequencies $\nu_\mu = \nu_\mathrm{p} + \mu f_{\mathrm{rep}}$ ($\mu$ is line index relative to the pump laser, $f_{\mathrm{rep}}$ is the pulse repetition rate). A second injection laser of frequency $\nu_\mathrm{inj}$, formally acting as a (weaker) second pump laser \cite{wildi2023SidebandInjectionLocking}, enables Kerr-nonlinear synchronisation so that $\nu_\mathrm{p}$ and $\nu_\mathrm{inj}$ are both part of the microcomb spectra (see \reffig{fig:concept}{b}). Effectively, this implements optical frequency division (OFD)~\cite{fortier2011GenerationUltrastableMicrowaves} where the frequency difference between the lasers is divided by an integer $N$ into the microcomb microwave repetition rate $f_\mathrm{rep}= |\nu_\mathrm{p}-\nu_\mathrm{inj}|/N$ (\reffig{fig:concept}{b}).  
Both pump and injection lasers are combined and delivered over fiber to separate microcombs, implying that all comb frequencies and, in particular, the repetition rates of both combs are synchronised. However, both microcombs are still subject to uncorrelated noise inducing relative timing jitter, notably fundamental thermorefractive noise \cite{kondratiev2018ThermorefractiveNoiseWhispering, huang2019ThermorefractiveNoiseSiliconnitridea}.

In a \textit{first} experiment, to investigate the synchronisation performance, we utilise a frequency domain technique, similar to the one developed for optical linewidth measurements in microcombs \cite{lei2022OpticalLinewidthSoliton}. It is based on the elastic tape model describing the phase noise PSDs of the frequency comb lines\cite{liehl2019DeterministicNonlinearTransformations, lei2022OpticalLinewidthSoliton} 
\begin{equation}
S_\mu(f) = S_\mathrm{fix}(f) + (\mu - \mu_{\mathrm{fix}})^2 S_{\mathrm{rep}}(f),
\label{eq:tapemodel}
\end{equation}
where $f$ is the noise frequency, $S_\mathrm{fix}$ represents the phase noise power spectral density (PSD) at the fix point at $\mu_{\mathrm{fix}}$ (not necessarily an integer), and  $S_{\mathrm{rep}}$ is the repetition rate phase noise PSD. The position of the fix point depends on the power levels and frequencies of the two continuous-wave lasers defining the comb's `centre of gravity'~\cite{wildi2023SidebandInjectionLocking}, and additional processes such as stimulated Raman scattering that can induce correlations between pump frequency and repetition rate noise~\cite{lei2022OpticalLinewidthSoliton}.
In our case, we assume both microcombs have approximately the same $\mu_{\mathrm{fix}}$, which we will justify later. This implies that Eq.~\ref{eq:tapemodel} can also describe the \textit{relative} phase noise PSDs between the microcombs. Thus, considering all PSDs in Eq.~\ref{eq:tapemodel} as relative noise quantities between the microcombs, we can retrieve the relative repetition rate phase noise $S_{\mathrm{rep}}^\mathrm{rel}(f)$ and relative timing jitter between the microcombs, based on measuring $S_\mu^\mathrm{rel}(f)$ in dependence of $\mu$ and $f$. 

\begin{figure*}[!t]
  \centering
  \includegraphics[width=\textwidth]{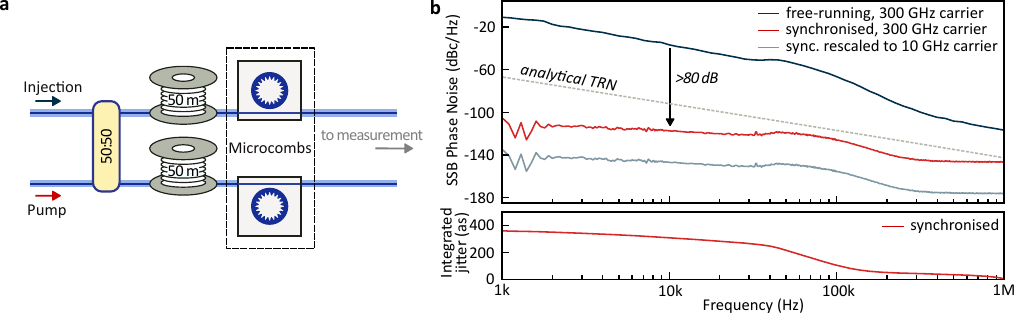}
  \caption{
    \textbf{Phase noise and timing jitter of synchronised 300~GHz microcombs.} 
    \textbf{a}, Experimental setup with 50 m of spooled single-mode fiber added in both arms before the chips.
    \textbf{b}, (top) Relative repetition rate single-sideband (SSB) phase-noise power spectral density in the free-running (dark blue) and synchronised (red) states at 300~GHz carrier. The dashed line indicates an analytic estimate\cite{kondratiev2018ThermorefractiveNoiseWhispering} of the fundamental thermorefractive noise (TRN) for the 300~GHz microresonator, and the gray trace shows the relative repetition rate noise rescaled to 10~GHz carrier. (bottom) The corresponding integrated residual timing jitter in the synchronised state. 
    }
  \label{fig:jitter}
\end{figure*}

The experimental setup for Kerr-nonlinear synchronisation and characterisation is shown in \reffig{fig:sync}{a}. It consists of three main blocks: the pump source, the microcombs, and the characterisation stage. We use two free-running, continuously tunable cw lasers. The main pump laser ($\sim 1560$~nm) is amplified to $\sim 630$~mW using an EDFA. The amplified spontaneous emission (ASE) is suppressed with a fiber Bragg grating (FBG) of $\sim 40$~GHz bandwidth centred at the pump wavelength. The pump is then combined with the injection laser ($\sim 1618$~nm) on a 50:50 fiber coupler. The two outputs of the coupler are used to drive two photonic crystal ring microresonators (PhCRs) on two different silicon nitride (Si$_3$N$_4$) integrated photonic chips. Both PhCRs have an intrinsic linewidth of $\sim 50$~MHz, a free-spectral range (FSR) of $\sim 300$~GHz (ring radius $\sim 75~\mu m$), and are critically coupled; the waveguide cross-section is $1.6\times0.8~\mu m^2$. Each PhCR features a periodic corrugation pattern along the inner wall with a spatial period of $\sim 422$~nm creating a mode-frequency splitting of the $\mu=0$ resonance of $\sim 250$~MHz, enabling deterministic excitation of single-soliton microcombs \cite{yu2021SpontaneousPulseFormation, ulanov2024SyntheticReflectionSelfinjectionlocked}. Optical input-output coupling to the photonic chips is implemented using ultra-high-numerical-aperture fibers with index-matching gel to suppress parasitic back-reflections, resulting in approximately 100~mW of on-chip pump power per chip. The microcombs with $f_\mathrm{rep}\approx300$~GHz are initiated by adjusting the detuning of the $\mu=0$ resonance with regard to the pump laser frequency $\nu_\mathrm{p}$ via chip-integrated electric microheaters. To investigate the relative timing jitter between the two microcombs, one of them is shifted in frequency by 80~MHz via an acousto-optic modulator (AOM; see \reffig{fig:sync}{a}), before they are recombined via a 50:50 fiber coupler. This ensures that beatnotes between the combs of lines with the same $\mu$ are shifted away from zero frequency, especially in the synchronised case.

After rejecting the pump line with an FBG, we obtain $\sim 2$~mW of total optical power. This signal is amplified with a second EDFA to 200~mW and passed through a tunable fiber-coupled bandpass filter (implemented using a commercial waveshaper), which can either transmit the entire spectrum or select individual comb lines with $\sim 25$~GHz bandwidth. The filtered signal is detected on a low-noise photodiode and analysed using an electrical spectrum analyser (ESA).

\begin{figure*}[!ht]
  \centering
  \includegraphics[width=\textwidth]{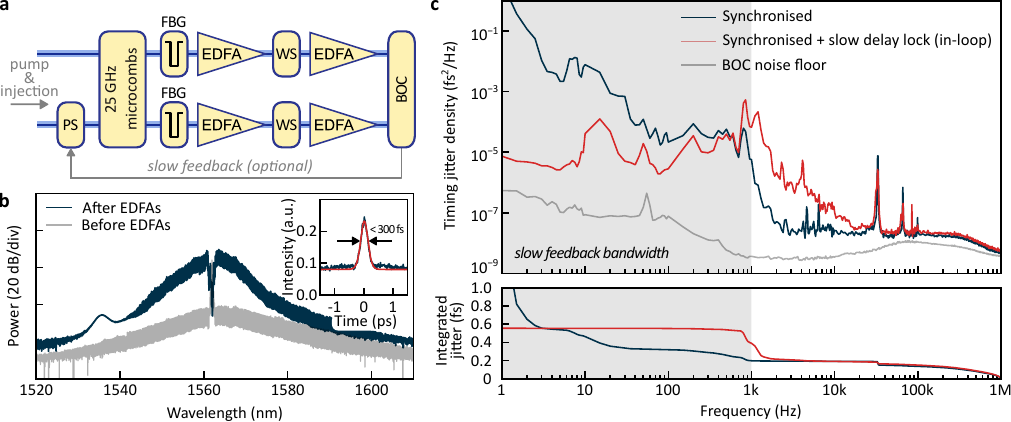}
  \caption{\textbf{Phase noise and timing jitter of synchronised 25~GHz microcombs.}
    \textbf{a}, Experimental setup with two 25~GHz microcomb. The balanced optical cross-correlation (BOC) method is used to characterise their relative timing-jitter. FBG - fiber Bragg grating, EDFA - erbium-doped fiber amplifier, WS - waveshaper, PS - piezo fiber stretcher.
    \textbf{b}, Optical spectrum of one of the 25~GHz microcombs out-coupled from the chip (gray) and after the second EDFA (blue) with the corresponding autocorrelation trace (inset; blue). The $\mathrm{sech}^2$ fit (inset; red) yields $\sim 300$~fs pulse duration.
    \textbf{c}, (top) Relative timing-jitter spectral density of two optically synchronised 25~GHz repetition-rate microcomb sources (blue). The red trace represents an in-loop measurement where the relative delay is locked via piezo fiber-stretcher feedback. The gray trace shows the BOC noise floor defined by the electric noise floor of the balanced photodetector. The gray shading indicates the frequency range where timing noise can be suppressed by feedback.
    }
    \label{fig:25GHz}
\end{figure*}

Initially, we characterise the microcombs in the free-running regime (injection laser off), where both microcombs operate with a repetition rate difference of $\Delta f_\mathrm{rep} \approx 5$~MHz. Their combined optical spectra are shown in \reffig{fig:sync}{b} (top panel). We sequentially center the bandpass filter on each pair of microcomb lines with $\mu \in [-4, 4]$ and measure with the ESA the single-sideband (SSB) phase noise $S_{\mu}^\mathrm{rel}(f)$ of the corresponding beatnote.
The results, shown in \reffig{fig:sync}{c}, indicate that the relative phase noise between comb lines increases with 
$|\mu|$, with the lowest noise observed at the $\pm 1$ sidebands (for better visibility we only show the traces for $\mu=1..4$). In this measurement, we are
only interested in the relative phase noise between the comb lines; the beatnote carrier frequency does not enter our considerations. From these measurements, we extract the phase noise at a fixed offset (10~kHz) and plot it as a function of line number $\mu$, revealing an almost ideal parabolic dependence on the mode number $\mu$ (see \reffig{fig:sync}{d}). This is consistent with the elastic tape model, where both combs are independently `breathing' around their common spectral fix point with $S_\mathrm{rep}(f)$. The fix point is here close to the pump laser at $\mu=0$ and additional relative fix point noise $S_\mathrm{fix}^\mathrm{rel}(f)$ that would manifest as a vertical offset of the parabola is negligible (see Supplementary Information (SI), Sec.~1, Fig.~S1).

Next, we set $\Delta f_\mathrm{rep}$ close to zero. This is readily possible as it can be continuously tuned via the microheaters across multiple tens of megahertz, while maintaining microcomb operation. To trigger Kerr-nonlinear synchronisation, we switch on the injection laser and tune it toward the $\mu'=23$ comb lines ($\approx 1618$~nm). When sufficiently close to the comb lines, the injection laser enters the sideband injection-locking range $\delta_\mathrm{lock} \propto \mu'^2 \sqrt{P_\mathrm{inj} P_{\mu'}}$, where $P_\mathrm{inj}$ and $P_{\mu'}$ are the injected power and the power of the comb line $\mu'$, respectively \cite{wildi2023SidebandInjectionLocking}. In our case, the sideband injection locking range $\delta_\mathrm{lock}$ exceeds 200~MHz, and the achievable locking bandwidth can be estimated to be of the same order\cite{sun2025MicrocavityKerrOptical}. Once locked, the $\mu'$ lines of both combs latch onto the injection laser, forcing their frequencies to coincide with the injection laser frequency, resulting in $\Delta f_\mathrm{rep}=0$. All beatnotes between comb line pairs are now at the AOM frequency $f_m=80$~MHz. 

In this synchronised state, we again measure the RF beatnote SSB phase noise (see \reffig{fig:sync}{c}) for comb sidebands $\mu \in \left[-6, 8 \right]$, revealing a drastically reduced relative phase noise, indicating tight synchronisation. Here, the injection laser plays two roles: (i) it provides a second frequency pin-point for both microcombs, which, through OFD, defines their repetition rates according to $f_\mathrm{rep} = |\nu_\mathrm{p} - \nu_\mathrm{inj}|/\mu'$ and forces $\Delta f_\mathrm{rep} = 0$; (ii) it suppresses absolute repetition rate phase noise via OFD \cite{sun2025MicrocavityKerrOptical}.
As before, we use the measured data to compute the phase noise at a given offset frequency for each measured pair of comb lines and plot the results as a function of the line number (see \reffig{fig:sync}{d}). The results again reveal a parabolic dependence, consistent with the elastic tape model, but this time with a shifted fix point $\mu_\mathrm{fix}$, that we assume is the same for both microcombs. This assumption is motivated by the fact that the PhCRs are of identical design and driven with nearly equal laser powers and justified by the fact that the data is well described by a single parabola. These observations confirm the validity of our approach. 

Before proceeding with our analysis, we introduce $50$~m of single-mode fiber (spooled) into both arms before the chips (see \reffig{fig:jitter}{a}) to emulate a realistic timing distribution scenario across 100~m distance.
In this configuration, we repeat the measurement of the phase noise to obtain $S_{\mu}^\mathrm{rel}(f)$ in the free-running and synchronised regimes. To explicitly derive $S_\mathrm{rep}^\mathrm{rel}(f)$ (and the timing jitter), we fit the parabolic $S_{\mu}^\mathrm{rel}(f)$ at each noise frequency $f$ (766 points along the $f$ axis). Based on Eq.~\ref{eq:tapemodel}, we determine the phase noise PSD $S_\mathrm{rep}^\mathrm{rel}(f)$ of the 300~GHz repetition rate signal, as shown in \reffig{fig:jitter}{b} (top): Kerr-nonlinear synchronisation results in a striking reduction of the phase noise compared to the free-running case by more than eight orders of magnitude at 10~kHz, to well below the thermorefractive noise level (for spectral dependence see SI, Sec.~2, Fig.~S2). Notably, this performance is achieved without any active feedback loops, only relying on the Kerr-nonlinear synchronisation by the two free-running cw lasers. For comparison with other oscillators, \reffig{fig:jitter}{b} also shows the corresponding spectrum rescaled to a 10~GHz carrier frequency. In \reffig{fig:jitter}{b} (bottom) we plot the integrated relative timing jitter between both combs, which is below 380~attoseconds when integrated from 1~kHz to 1~MHz. A comparative measurement with and without the additional 2×50 m fiber shows no measurable impact on phase noise or timing jitter, indicating that longer links are feasible (see SI, Fig.~S1). The present limitation arises from unshielded fiber components rather than the spooled fiber coils. Fiber length may ultimately be constrained by nonlinear effects such as stimulated Brillouin scattering of the CW pumps, unless mitigated (e.g., with isolators). If active stabilisation is required, the fiber delay must remain shorter than the stabilisation time scale.

In a \textit{second} experiment, we further extend the Kerr-nonlinear synchronisation to an electrically detectable repetition rate of 25~GHz. This addresses applications complementary to those of the 300~GHz system, and also enables advanced modulation schemes for robust optical time transfer~\cite{caldwell2025HighprecisionOpticalTime}. The lower repetition rate enables us to use BOC~\cite{schibli2003AttosecondActiveSynchronization, xin2018UltrapreciseTimingSynchronization} to independently validate the low-timing jitter performance. The setup is shown in \reffig{fig:25GHz}{a}; note, we do not insert 2x50~m of fiber as we did not observe a noticeable impact on the timing jitter before (see SI, Fig.~S1). We now operate with 25~GHz FSR racetrack resonators, which are fabricated on the same silicon nitride platform. The resonators are critically coupled, the waveguide cross-section is $2.0\times0.8~\upmu$m$^2$ and the intrinsic linewidth is $25$~MHz. As their 300~GHz counterparts, they feature a periodic corrugation pattern to enable deterministic excitation of single soliton microcombs. After rejection of the 260~mW pump laser via FBGs, the microcombs are amplified via EDFAs. The amplifier dispersion is compensated via waveshapers, yielding trains of $~300$~fs pulses (see \reffig{fig:25GHz}{b}), each with an average power of $\sim 70$~mW. 

The results of the BOC measurement are shown in \reffig{fig:25GHz}{c}. Kerr-nonlinear synchronisation results in sub-200~as relative integrated timing jitter when integrated from 1~kHz to 1~MHz. When integrated down to 1~Hz, we observe approximately 1~fs of integrated timing jitter, entirely limited in our case by the effective length drift of the unstabilised fibers. 
In this context, we note that a fiber link can be readily stabilised via a simple piezo actuator on time-scales of 1~ms (below 1~kHz). 
To illustrate this, we perform a stabilisation experiment, feeding the BOC signal back to a fiber stretcher in the fiber link to one of the combs. In this way, the in-loop timing jitter remains constant below 1~kHz. This does not confirm the out-of-loop timing performance, however, it demonstrates the actuator effectiveness below 1~kHz. We note that the link tracking signal for stabilisation can also be derived directly from the two-tone reference~\cite{roslund2025OpticalTwotoneTime}, in our case provided by the pump and injection lasers, without requiring BOC. Together, these experiments demonstrate the capabilities needed to implement all-optical synchronisation in multi-node optical networks.

\subsection*{Conclusion}

In conclusion, we demonstrate the first attosecond-level synchronisation of microcomb oscillators, at both 25~GHz and 300~GHz repetition rate. The synchronisation is achieved by transmitting a main pump laser and a second injection laser over fiber to the two microcombs. Kerr-nonlinear synchronisation reduces the relative phase noise and timing jitter between both oscillators by more than 80 dB resulting in a remarkably low integrated relative timing jitter of less than 400~as. While in our case the lasers were free-running, they could be stabilised to a low noise reference cavity, or derived from a low-noise mode-locked laser, enabling precision synchronisation between complementary laser technologies. These results could enable precision timing across large networks in attosecond experiments, radio-telescopes, geodesy, or in increasingly demanding and emerging technological applications such as data centers and large-scale computing facilities \cite{clark2020SynchronousSubnanosecondClock, xue2022NanosecondOpticalSwitching, gonzalez2022OpticallyConnectedMemory}, quantum-secure communication and quantum-information\cite{clivati2022CoherentPhaseTransfer}. 
Ultimately, our results may lead to chip-integrated attosecond photonics.

\small
\paragraph{Funding.}
\small
This project has received funding from the European Research Council (ERC) under the EU's Horizon 2020 research and innovation program (grant agreement No 853564), from the EU's Horizon 2020 research and innovation program (grant agreement No 101137000) and through the Helmholtz Young Investigators Group VH-NG-1404; the work was supported through the Maxwell computational resources operated at DESY.

\printbibliography
\end{refsection}

\newpage
\begin{refsection}

\setcounter{equation}{0}
\setcounter{figure}{0}
\renewcommand{\thefigure}{S\arabic{figure}}









\title{Supplementary Information - \\Attosecond-level synchronisation of chip-integrated oscillators}



\onecolumn
\maketitle

\section{Investigation of the relative repetition rate $S^\mathrm{rel}_\mathrm{rep}(f)$ and offset noise $S^\mathrm{rel}_\mathrm{fix}(f)$}
\label{sec:100mfiber}

\begin{figure*}[h]
  \centering
  \includegraphics[width=0.75\linewidth]{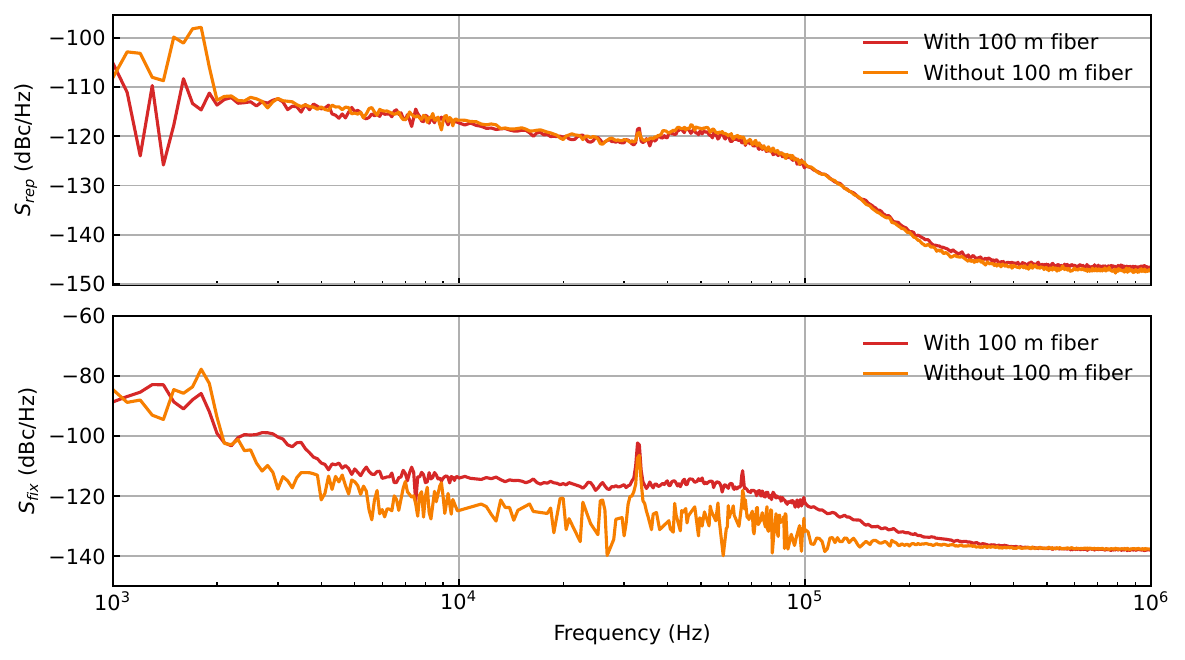}
  \caption{Relative repetition rate (top) and offset (bottom) single-sideband phase-noise power spectral densities $S^\mathrm{rel}_\mathrm{rep}(f)$  and $S^\mathrm{rel}_\mathrm{fix}(f)$ reconstructed via fitting at each Fourier frequency with (red) and without (orange) a 100~m fiber link.
  }
  \label{sifig:fig1}
\end{figure*}

In the synchronised state, we record the IQ data of the inter-comb beatnote for the sidebands $\mu \in \left[-6, 8 \right]$, both with and without a 100~m spooled fiber link between the two microcombs (cf. main text). We use these data and perform a windowed Welch transform to reconstruct the relative single-sideband phase noise of the corresponding sidebands. To derive the relative repetition-rate (and hence timing-jitter) noise, we fit a parabolic curve at each analysis frequency $f$, following the procedure explained in the main text. In this approach, the parabolic scaling directly yields the relative repetition-rate noise spectrum $S^\mathrm{rel}_\mathrm{rep}(f)$, while the parabola offset corresponds to $S^\mathrm{rel}_\mathrm{fix}(f)$. The obtained data comparing the cases with and without the fiber link are presented in \reffig{sifig:fig1}{}.

These experiments show that a longer fiber results in a slightly increased $S^\mathrm{rel}_\mathrm{fix}(f)$, while the impact on $S^\mathrm{rel}_\mathrm{rep}(f)$ is not noticeable for frequencies down to 1~kHz. 

\newpage
\section{Sideband injection locking relative noise reduction}

\begin{figure*}[h]
  \centering
  \includegraphics[width=0.75\linewidth]{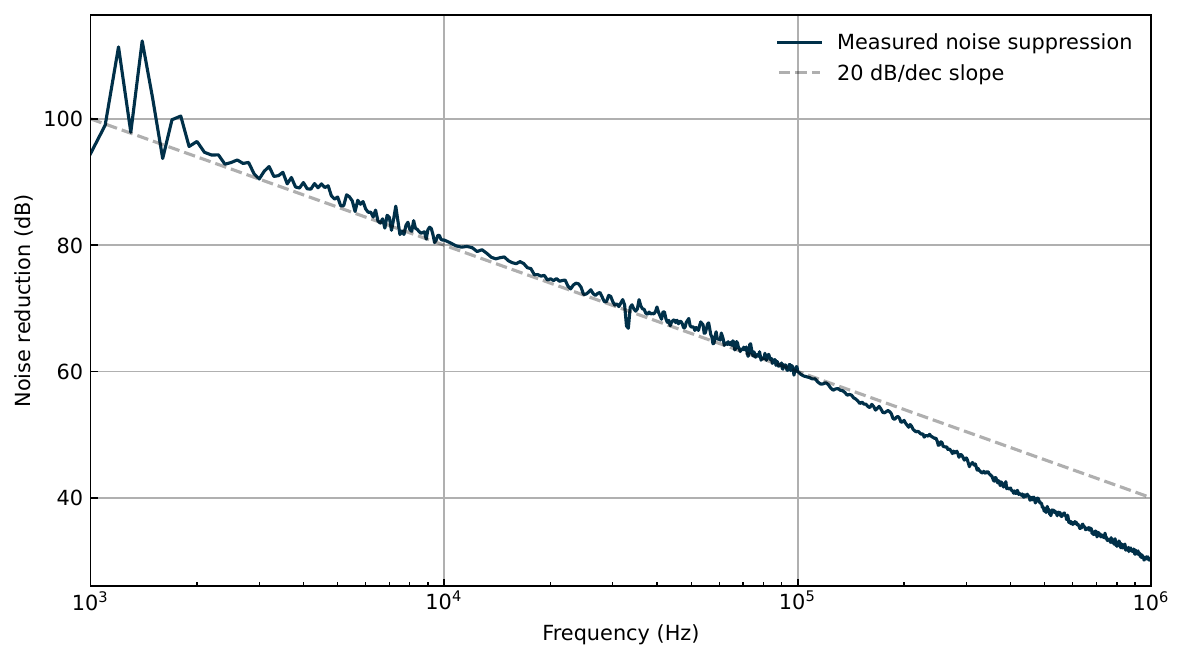}
  \caption{Relative single-sideband phase-noise suppression as a function of frequency inferred from Fig.~3b (cf. main text).
  }
  \label{sifig:fig2}
\end{figure*}

In \reffig{sifig:fig2}{} we show the relative single-sideband phase-noise suppression as a function of frequency, as inferred from Fig.~3b (cf. main text). In the frequency range from 1~kHz to 100~kHz, where differential fiber drifts are negligible, and the signal remains well above the white-noise floor, the noise reduction closely follows a 20~dB-per-decade trend, as predicted by theory \cite{sun2025MicrocavityKerrOptical}.

\printbibliography


\end{refsection}

\end{document}